\documentclass[5p]{elsarticle}
\pdfoutput=1
\usepackage[english]{babel}
\usepackage[latin1]{inputenc}
\usepackage{amsfonts,amsbsy,bm,euscript,mathrsfs}
\usepackage{amssymb,stmaryrd,faktor,slashed}
\usepackage[x11names]{xcolor}
\usepackage[tbtags]{amsmath}
\usepackage[bookmarks=true,colorlinks=true,linkcolor=black,citecolor=black,urlcolor=black,bookmarksnumbered]{hyperref}

\newcommand{\foot}[1]{\footnote{#1\vspace{2pt}}}

\def\ads{{\rm AdS}_5\times {\rm S}^5}

\title{Almost abelian twists and AdS/CFT\\
\vspace{-50pt}
\begin{flushright} \small \texttt{HU-EP-16/35\\ HU-MATH-16/19}\end{flushright}}
\author{Stijn J. van Tongeren}
\ead{svantongeren@physik.hu-berlin.de}
\address{Institut f\"ur Mathematik und Institut f\"ur Physik, Humboldt-Universit\"at zu Berlin, \\ IRIS Geb\"aude, Zum Grossen Windkanal 6, 12489 Berlin, Germany}

\begin{document}
%%%%%%%%%%%%%%%%%%%%%%%%%%%%%%%%%%%%%%

\begin{abstract}
A large class of the recently found unimodular nonabelian homogeneous Yang-Baxter deformations of the $\ads$ superstring can be realized as sequences of noncommuting TsT transformations. I show that many of them are duals to various noncommutative versions of supersymmetric Yang-Mills theory, structurally determined directly in terms of the associated $r$ matrices, in line with previous expectations in the literature.
\end{abstract}


\maketitle

%%%%%%%%%%%%%%%%%%%%%%%%%%%%%%%%%%%%%%
\section{Introduction}\label{sec:intro}
%%%%%%%%%%%%%%%%%%%%%%%%%%%%%%%%%%%%%%

Integrable models arise throughout physics as useful case studies balancing complexity and solvability. In the context of the AdS/CFT correspondence  \cite{Maldacena:1997re}, it is possible to perform detailed tests of this conjecture, and get remarkable insight into four dimensional planar gauge theory, based on the integrability of the $\ads$ superstring and its dual, planar $\mathcal{N}=4$ supersymmetric Yang-Mills theory (SYM).\foot{For reviews see e.g. \cite{Arutyunov:2009ga,Beisert:2010jr,Bombardelli:2016rwb}.} Beyond appearing in further lower dimensional examples of AdS/CFT, integrability is also preserved by certain deformations of this canonical duality, such as the $\beta$ deformation of SYM dual to strings on the Lunin-Maldacena background \cite{Lunin:2005jy,Frolov:2005ty,Frolov:2005dj}. Rather than looking for integrability in specific AdS/CFT dual pairs however, recent years have instead seen a focus on finding integrable deformations of just the $\ads$ string $\sigma$ model. Many such models can be generated as Yang-Baxter (YB) deformations \cite{Klimcik:2002zj,Klimcik:2008eq} of the string $\sigma$ model, introduced in \cite{Delduc:2013qra}.\footnote{A second type of deformation gives the $\lambda$ model \cite{Sfetsos:2013wia,Hollowood:2014qma,Demulder:2015lva} that can be naturally viewed as a deformation of the nonabelian $T$ dual of the $\sigma$ model.} The original YB deformation gives the $\eta$ model, which algebraically corresponds to quantum deforming the symmetry algebra of the string \cite{Delduc:2013qra,Delduc:2014kha}. This deformation is based on an inhomogeneous $r$ matrix solving the modified classical Yang-Baxter equation (mCYBE), but it can be generalized to homogeneous $r$ matrices solving the regular classical Yang-Baxter equation (CYBE) \cite{Kawaguchi:2014qwa}. This leads to Drinfeld twisted symmetry \cite{vanTongeren:2015uha}. Being manifestly integrable, the important questions are now: are the resulting models still string theories, and if so, do they have an AdS/CFT interpretation?

The first of these question has recently been answered in general. Namely, the resulting model is conformally invariant (represents a string) at one loop if and only if the associated $r$ matrix is unimodular \cite{Borsato:2016ose}. At present no unimodular solution of the mCYBE is known, and indeed the $\eta$ model is not a string \cite{Arutyunov:2015qva}, nor are two closely related formulations with inequivalent backgrounds \cite{Hoare:2016ibq}. In terms of the CYBE, abelian $r$ matrices correspond to TsT transformations (Melvin twists) \cite{vanTongeren:2015soa,Osten:2016dvf}, hence give string theories, and are trivially unimodular. The other previously studied case of bosonic jordanian $r$ matrices does not give string theories \cite{Kyono:2016jqy,Hoare:2016hwh,Orlando:2016qqu} -- many of the models are in fact related to inhomogeneous $\eta$-type models by singular boosts \cite{Hoare:2016hwh} -- and indeed is not unimodular.\footnote{The backgrounds of all YB models solve modified supergravity equations \cite{Arutyunov:2015mqj}, as required by $\kappa$-symmetry \cite{Wulff:2016tju}.} It turns out that the symmetry algebra of $\mathrm{AdS}_5$, $\mathfrak{so}(4,2)$, admits 17 inequivalent homogeneous nonabelian unimodular $r$ matrices of rank four, and at least one of rank six \cite{Borsato:2016ose}. Including an 18th one that cannot be realized just within $\mathfrak{so}(4,2)$, these can all be extended using generators of $\mathfrak{so}(6)$, leading to many more options for the full bosonic symmetry algebra of the string. Their nonabelian structure always resides in $\mathfrak{so}(4,2)$ however. The associated deformations correspond to nonabelian T duality in string theory \cite{Hoare:2016wsk,Borsato:2016pas}. This leaves the question whether the resulting models have an AdS/CFT interpretation, and if so, what their duals are.

Before these new unimodular models were known, I conjectured that homogeneous YB deformations with a string theory interpretation are in general dual to noncommutative (NC) field theories \cite{vanTongeren:2015uha}. In this picture, the twisted symmetry of these YB models is implemented on the field theory side by introducing $\star$ products in spacetime or (super)field space. This provides a uniform picture for many known AdS/CFT pairs such as the $\beta$ deformation dual to the Lunin-Maldacena background and canonical NC SYM and its gravity dual \cite{Hashimoto:1999ut,Maldacena:1999mh}, which can be realized as abelian YB deformations \cite{Matsumoto:2014nra,Matsumoto:2014gwa,Kyono:2016jqy}. Here I will briefly show that, as expected, this picture indeed applies to many of the new unimodular models.

Of the new deformations found in \cite{Borsato:2016ose}, I will consider those that: 1) have an ``almost abelian'' structure that allows them to be interpreted as sequences of noncommuting TsT transformations in string theory,\footnote{TsT transformations, standing for T duality - shift - T duality as discussed below, can be viewed as a special case of nonabelian $T$ duality \cite{Hoare:2016wsk}.} and 2) are generated by elements of $\mathfrak{iso}(3,1) \subset \mathfrak{so}(4,2)$, meaning part of the symmetries of ten dimensional flat space. This covers 12 out of the 17 possible $\mathfrak{so}(4,2)$ rank four deformations, and the rank six deformation, given in \cite{Borsato:2016ose}. This structure naturally suggests a deformation of flat space to place branes in, where an appropriate low energy limit gives either an open string picture for NC field theories of the desired type in the spirit of \cite{Schomerus:1999ug,Seiberg:1999vs}, or as its dual precisely the associated YB deformation of $\ads$. In particular, on the field theory side I show that the NC parameter $\theta$ is simply the $r$ matrix, providing an explicit match with the expected $\star$ product. As issues are known to arise in taking low energy field theory limits when considering time-space noncommutativity (electric $B$ fields) \cite{Seiberg:2000ms}, combinations involving such cases should be excluded however. Up to this restriction, the resulting pairs of theories are dual in the sense of AdS/CFT, at least in supersymmetric cases. I illustrate this construction explicitly for two examples.

This picture can be readily extended to include nonabelian $r$ matrices with generators of $\mathfrak{so}(6)$, also acting naturally on flat space. From the TsT picture one expects to find dipole deformations, see e.g. \cite{Dasgupta:2001zu}, of the above types of NC SYM. This is precisely in line with the general twist proposal of \cite{vanTongeren:2015uha}, cf. footnote \ref{footnote:AATsT=rdef} below.

In the next section I briefly recall the $\ads$ string $\sigma$ model and its YB deformation, and the type of $r$ matrices to be considered, with two explicit nonabelian examples that can be realized via noncommuting TsT transformations. In section \ref{sec:adscft} I discuss the AdS/CFT interpretation of these almost abelian twisted models, with two explicit examples. I conclude with open questions.

\section{Yang-Baxter $\sigma$ models}
\label{sec:modelintro}

Homogeneous YB deformations of the $\mathrm{AdS}_5 \times \mathrm{S}^5$ superstring action are of the form \cite{Delduc:2013qra,Kawaguchi:2014qwa}\foot{Here $T$ is the would-be effective string tension, $h$ is the world sheet metric, $\epsilon^{\tau\sigma}=1$, $A_\alpha = g^{-1} \partial_\alpha g$ with $g\in \mathrm{PSU}(2,2|4)$, $\mathrm{sTr}$ denotes the supertrace, and $d_\pm = \pm P_1 + 2 P_2 \mp P_3$ where the $P_i$ are the projectors onto the $i$th $\mathbb{Z}_4$ graded components of the semi-symmetric space $\mathrm{PSU}(2,2|4)/(\mathrm{SO}(4,1)\times \mathrm{SO}(5))$ (super $\mathrm{AdS}_5 \times \mathrm{S}^5$).}
\begin{equation}
\label{eq:defaction}
S = -\tfrac{T}{2} \int d\tau d\sigma \tfrac{1}{2}(\sqrt{h} h^{\alpha \beta} -\epsilon^{\alpha \beta}) \mathrm{sTr} (A_\alpha d_+ J_\beta),
\end{equation}
where $J=(1-\eta R_g \circ d_+)^{-1}(A)$ with $R_g(X)=g^{-1} R(g Xg^{-1}) g$.  The operator $R$ is a linear map from $\mathfrak{g}=\mathfrak{psu}(2,2|4)$ to itself. $\eta=0$ ($R=0$) corresponds to the undeformed $\mathrm{AdS}_5 \times \mathrm{S}^5$ superstring action of \cite{Metsaev:1998it}. Now, provided $R$ is antisymmetric, $\mathrm{sTr}(R(m) n) = -\mathrm{sTr}(m R(n))$,
%\begin{equation}
%\mathrm{sTr}(R(m) n) = -\mathrm{sTr}(m R(n)),
%\end{equation}
and satisfies the classical Yang-Baxter equation (CYBE)
\begin{equation}
\label{eq:CYBEOP}
[R(m),R(n)] - R([R(m),n] + [m,R(n)])=0,
\end{equation}
this deformed model is classically integrable and has a form of $\kappa$ symmetry.

These $R$ operators are related to $r$ matrices via a nondegenerate bilinear form on $\mathfrak{psu}(2,2|4)$, induced by the Killing form of $\mathfrak{su}(2,2|4)$. Using a matrix representation of $\mathfrak{su}(2,2|4)$
\begin{equation}
R(m) = \mathrm{sTr}_2(r (1\otimes m)),
\end{equation}
with
\begin{equation}
r = \sum_{i,j} \alpha_{ij} t^i \wedge t^j  \in \mathfrak{g} \otimes \mathfrak{g},
\end{equation}
where the $t^i$ generate $\mathfrak{g}$, $\alpha_{ij} \in \mathbb{R}$, $a \wedge b = a\otimes b - b \otimes a$, and $\mbox{sTr}_2$ denotes the supertrace over the second space in the tensor product. Equation \eqref{eq:CYBEOP} translates to equation \eqref{eq:CYBEMAT} in matrix form. I will refer to both the operator $R$ and the matrix $r$ as the $r$ matrix.

A YB deformation preserves symmetries generated by those $t \in \mathfrak{g}$ for which
\begin{equation}
R([t,x]) = [t,R(x)] \ \forall \, x \in \mathfrak{g}.
\end{equation}
These are symmetries of the $r$ matrix in the sense that
\begin{equation}
(\mbox{ad}_t \otimes 1 + 1 \otimes \mbox{ad}_t) \, r = 0,
\end{equation}
where $\mbox{ad}$ denotes the adjoint action. The remaining symmetry is deformed by the Drinfeld twist associated to $r$ \cite{vanTongeren:2015uha}.

Given an $r$ matrix and a coset parametrization $g$, inverting $1-\eta R_g \circ d_+$ and comparing to the standard Green-Schwarz superstring action gives an explicit background for the $\sigma$ model.

\subsection*{Almost abelian $r$ matrices}

I will consider rank\footnote{The rank is the number of independent algebra elements used to construct it.} four $r$ matrices of the form \cite{Borsato:2016ose}
\begin{equation}
r= a \wedge b + c \wedge d,
\end{equation}
where the generators $a$, $b$, $c$ and $d$ generate a unimodular quasifrobenius subalgebra of $\mathfrak{so}(4,2)$ \cite{Ovando:2006}. Of the possible nonabelian cases I consider
\begin{itemize}
\item $\mathfrak{h}_3 \oplus \mathbb{R}$
 \begin{equation}
 \label{eq:comm1}
 [c,a] = b
 \end{equation}
 \item $\mathfrak{r}^\prime_{3,0}\oplus \mathbb{R}$
 \begin{equation}
 \label{eq:comm3}
 [c,a]=-b, \quad [c,b]=a
 \end{equation}
\item $\mathfrak{r}_{3,-1}\oplus \mathbb{R}$
 \begin{equation}
 \label{eq:comm2}
  [c,a]=a, \quad [c,b]=-b
 \end{equation}
\end{itemize}
where I have indicated their defining Lie brackets. For all these, $[a,b]=0$, and $d$ is central. Due to above relations, $c$ and $d$ are symmetries of the $r$ matrix.\footnote{The $r$ matrices associated to the fourth type of subalgebra in \cite{Borsato:2016ose}, $\mathfrak{n}_4$, do not have this property.} I will refer to these $r$ matrices as ``almost abelian'', compared to abelian $r$ matrices constructed out of a set of commuting generators. It will be useful to split such $r$ in two abelian pieces $\hat{r}$ and $\bar{r}$ as
\begin{equation}
r = \hat{r} + \bar{r}, \quad \hat{r} = a \wedge b, \quad \bar{r} = c \wedge d.
\end{equation}
The fact that $\bar{r}$ is built out of generators of symmetries of $\hat{r}$ is also referred to as $\bar{r}$ being subordinate to $\hat{r}$, see e.g. \cite{Borowiec:2008se}. The rank six $r$ matrix given in \cite{Borsato:2016ose} is also of this form, i.e.
\begin{equation}
r = \hat{r} + \bar{r} + \tilde{r},
\end{equation}
where $\tilde{r}$ is subordinate to $\hat{r}$ and $\bar{r}$, and $\bar{r}$ is subordinate to $\hat{r}$, and all pieces are abelian. This structure makes it possible to directly construct the associated twists.

\subsection*{Examples}

Consider\footnote{I use $1/2$ times the $r$ matrices used in \cite{Borsato:2016ose}, up to signs. Moreover, for dimensional reasons one might want to formally insist on separate deformation parameters for the separate terms. They can be set numerically equal by an $\mathfrak{so}(4,2)$  automorphism however.}
\begin{equation}
r_1 = 2 m_{+3} \wedge p_+ + \tfrac{1}{2} p_2 \wedge p_3 = \hat{r}_1 + \bar{r}_1,
\end{equation}
and
\begin{equation}
r_2 = \tfrac{1}{2} p_1 \wedge p_2 +  \tfrac{1}{2} m_{12} \wedge p_3 = \hat{r}_2 + \bar{r}_2,
\end{equation}
where the $p$ and $m$ denote translation and Lorentz generators of $\mathfrak{so}(4,2)$ respectively (see \ref{app:confalg}), and I use light cone coordinates $x^\pm = x^0 \pm x^1$. This $r_1$ and $r_2$ are examples of $r$ matrices associated to $\mathfrak{h}_3 \oplus \mathbb{R}$ and $\mathfrak{r}_{3,0}^\prime\oplus \mathbb{R}$ as given above respectively.\footnote{As $\mathfrak{r}_{3,-1}\oplus \mathbb{R}$
is an analytic continuation of $\mathfrak{r}^\prime_{3,0}\oplus \mathbb{R}$ and this carries through the derivation, I will not consider an explicit example in this class. As discussed below, this formal continuation can have important consequences in the context of AdS/CFT however.}

%Taking coset parameterizations $g_1$ and $g_2$ as parametrizations of $\mathrm{AdS}_5$ for the $r_1$ and $r_2$ deformation respectively,
%\begin{align}
%g_1 = e^{x^\mu p_\mu} e^{\log z \, D},\qquad \qquad g_2 = e^{\xi M_{12}} e^{\rho  p_1 + x^0 p_0 + x^3 p_3} e^{\log z\, D},
%\end{align}
The background of the $\sigma$ model associated to $r_1$ is \cite{Borsato:2016ose}
\begin{align}
ds_1^2&=\frac{(dx^2)^2+(dx^3)^2+\eta ^2 z^{-4} { x^-} d { x^-} (2 d { x^2}- { x^-}
   d { x^-})}{z^2+\eta ^2/z^2}\nonumber \\
& \quad +\frac{-d { x^-} d { x^+}+dz^2}{z^2}+d\Omega_5^2, \label{eq:backgroundr1}\\
B_1& =\frac{\eta   ( dx^2-{ x^-} d { x^-}) \wedge d x^3}{ \left(\eta ^2+z^4\right)},\nonumber
\end{align}
where $d \Omega_5^2$ denotes the metric on $S^5$ and $x^\mu,z$ are Poincar\'e coordinates for $\mathrm{AdS}_5$. The background associated to $r_2$ is
\begin{align}
ds_2^2 & = \frac{z^4(d\rho^2 +\rho^2d\xi^2+ (d {x^3})^2) +\eta ^2 (\rho d \rho - dx^3)^2}{z^6 + \eta^2 z^2 (\rho^2+1)} \nonumber\\
& \quad +\frac{dz^2-(d {x^0})^2}{z^2}+d\Omega_5^2,\label{eq:backgroundr2}\\
B_2 & =\frac{\eta \,  \rho \, d\xi \wedge ( d\rho - \rho  d {x^3})}{z^4+\eta ^2 \left(\rho^2+1\right)}.\nonumber
\end{align}
where $\rho$ and $\xi$ are the radial and angular coordinate in the $x_1,x_2$ plane respectively, i.e. $\xi = \arctan x_2/x_1$. Beyond their existence, I will not need the dilaton and RR fields that complete these to full string backgrounds.

Importantly, these backgrounds can be realized by sequences of noncommuting TsT transformations \cite{Borsato:2016ose}. Denoting $T$ duality along an isometry coordinate $x$ by $T_x$, and the dualized coordinate by $\tilde{x}$, the first background is obtained from undeformed $\ads$ by the sequence
\begin{equation}
\label{eq:TsTr1}
T_\psi,\ w^+\to w^+\hspace{-2pt}-\eta \,\tilde{\psi},\ T_{\tilde\psi},\quad
T_{x^2},\ x^3\to x^3-\eta \,\tilde{x}^2,\ T_{\tilde{x}^2},
\end{equation}
where the first TsT transformation corresponding to $\hat{r}_1$ uses coordinates
\begin{equation}
x^+ = 2(\psi^2 w^- + w^+),\quad
x^- = 2w^-,\quad
x_3 = -2\psi w^-.
\end{equation}
The second background corresponds to
\begin{equation}
\label{eq:TsTr2}
T_{x^1},\ x^2\to x^2-\eta \,\tilde{x}^1,\ T_{\tilde{x}^1},\qquad
T_{\xi},\ x^3\to x^3-\eta \,\tilde{\xi},\ T_{\tilde{\xi}}.
\end{equation}
Similar considerations apply to any $r$ matrix of the above types \cite{Borsato:2016ose}, see also footnote \ref{footnote:AATsT=rdef} below. An abelian building block such as $\hat{r} = \tfrac{1}{2} a \wedge b$ is associated to a TsT transformation in $(y^{(a)},y^{(b)})$, meaning dualization in $y^{(a)}$ and shifting $y^{(b)}$, where $y^{(z)}$ denotes the coordinate dual to $z$ \cite{vanTongeren:2015soa}.\footnote{For instance, in terms of the coordinates above, $r_1 = \tfrac{1}{2}(\partial_\psi \wedge \partial_{w^+}+\partial_2 \wedge \partial_3)$ and $r_2 = \tfrac{1}{2}(\partial_1 \wedge \partial_2 + \partial_\xi \wedge \partial_3)$.} The sequence of the TsT transformations is determined by the subordinate structure. Let me now discuss the AdS/CFT interpretation of these almost abelian deformed strings.

%To elaborate, note that e.g. $\tilde{r}_1 = 2 M_{+3} \wedge p_+ + \alpha p_2 \wedge p_3 $ for arbitrary $\alpha$ is perfectly perfectly equivalent to $r_1$ as the two are related by an automorphism of the conformal algebra ($p \rightarrow a p, k\rightarrow a^{-1} k$). Setting $\alpha =0$ in $\tilde{r}_1$, however, gives the purely abelian $\hat{r}_1$, whose associated deformation is equivalent to a TsT transformation \cite{vanTongeren:2015soa,Osten:2016dvf}. Since $p_2$ and $p_3$ remain isometries of the background, we can subsequently apply the TsT transformation associated to $\bar{r}_1$, and this turns out to give the full background. Similar comments apply to any $r$ matrix of the types we are considering \cite{Borsato:2016ose}.

\section{AdS/CFT}
\label{sec:adscft}

Since the symmetries of these deformed strings are Drinfeld twisted, their hypothetical AdS/CFT duals should be able to realize twisted symmetry. This naturally leads to NC field theory, see e.g. the reviews \cite{Szabo:2006wx,Dimitrijevic:2014dxa}. Briefly, on the field theory side $\mathfrak{so}(4,2)$ is a spacetime symmetry, acting on the algebra of functions (fields) on Minkowski space. To define Drinfeld twisted $\mathfrak{so}(4,2)$ symmetry one needs to work with a different module, a deformed algebra of functions. Indeed, $F \in \mathcal{U}(\mathfrak{g}) \otimes \mathcal{U}(\mathfrak{g})$ can be used to define a twisted product between functions
\begin{equation}
f\star g \equiv \mu \circ F^{-1} (f \otimes g),
\end{equation}
where we understand $\mathfrak{so}(4,2)$ to be realized in terms of differential operators, see \ref{app:confalg}, and $\mu(f\otimes g) = fg$ denotes formal multiplication. For more details on twists, see \ref{app:Drinfeldtwist}. Taking $f$ and $g$ to be the coordinate functions on Minkowski space, this gives the basic NC structure of the theory as
\begin{equation}
[x^\mu \stackrel{\star}{,} x^\nu] \equiv x^\mu \star x^\nu - x^\nu \star x^\mu.
\end{equation}
For almost abelian $r$ matrices,
\begin{equation}
F = e^{i \eta \bar{r}} e^{i \eta \hat{r}}
\end{equation}
is an associated twist, matching the picture of the associated deformations being equivalent to a TsT transformation associated to $\hat{r}$, followed by the one for $\bar{r}$. For the rank six case we simply add a third term corresponding to $\tilde{r}$ to $F$, respectively apply the corresponding TsT transformation to the background. With this twist the formula above simplifies to
\begin{equation}
[y^\mu \stackrel{\star}{,} y^\nu] = -2 i \eta \mu(r(y^\mu \otimes y^\nu)) = -2 i \eta a^{\mu\nu},
\end{equation}
for all rank four $r$ matrices, in a set of appropriate coordinates $y$ where
\begin{equation}
\label{eq:rdiffop}
r = a^{\mu\nu}(y) \partial_{y^\mu} \wedge \partial_{y^\nu}.
\end{equation}
Namely, since the generators $a$, $b$, and $d$ of the $r$ matrix commute, it is possible to choose coordinates $(y^{(a)},y^{(b)},y^{(d)},\tilde{y})$ on $\mathbb{R}^{1,3}$ such that $z \sim \partial_{y^{(z)}}$. Then the remaining generator $c \sim y^{(b)} \partial_{y^{(a)}}$ for $\mathfrak{h}_3 \oplus \mathbb{R}$, $c \sim y^{(a)} \partial_{y^{(b)}} - y^{(b)} \partial_{y^{(a)}}$ for $\mathfrak{r}^\prime_{3,0}\oplus \mathbb{R}$, and $c \sim y^{(a)} \partial_{y^{(a)}} - y^{(b)} \partial_{y^{(b)}}$ for $\mathfrak{r}_{3,-1}\oplus \mathbb{R}$. In these coordinates $r^n(y^\mu \otimes y^\nu)$ vanishes for $n>1$ as $r$ lowers the polynomial order of the product of functions it acts on. This is not the case for e.g. $\tilde{r} = m_{01} \wedge m_{23}$ in cartesian coordinates -- part of the rank six $r = p_0 \wedge p_1 + p_2 \wedge p_3 + m_{01} \wedge m_{23}$ \cite{Borsato:2016ose} -- where the full twist is needed.

The twists for $r_1$ and $r_2$ for instance give nonvanishing
\begin{equation}
\label{eq:NC1}
[x^+ \stackrel{\star}{,} x^3]_1 =  -2 i \eta x^-, \qquad [x^2 \stackrel{\star}{,} x^3]_1  =  -i \eta,
\end{equation}
and
\begin{equation}
\label{eq:NC2}
\begin{aligned}
{\color{white} ?} [x^1 \stackrel{\star}{,} x^3]_2 = i \eta x^2,  \quad [x^2 \stackrel{\star}{,} x^3]_2 &=  -i \eta x^1,\\  \hspace{30pt} [x^1 \stackrel{\star}{,} x^2]_2 =  i \eta.
\end{aligned}
\end{equation}

In a conventional picture of the AdS/CFT correspondence, this structure should arise out of the low energy limit of open strings stretching between D3 branes, while the near horizon low energy limit of the same configuration should give the backgrounds of the corresponding deformed $\sigma$ models. To get a NC field theory in this picture, the background in which the branes are placed needs to be deformed. Restricting to $r$ matrices which only involve generators of $\mathfrak{iso}(3,1)$ as acting on $\mathbb{R}^{1,3} \subset \mathbb{R}^{1,9}$, finding such deformed backgrounds is not complicated.\footnote{This takes out $r_5$ and $r_6$ in \cite{Borsato:2016ose}, which involve generators outside of any one $\mathfrak{iso}(3,1)$ subalgebra. Any $\mathfrak{iso}(3,1)$ $r$ matrix that is connected by an inner automorphism to the ``$\mathbb{R}^{1,3}$'' $\mathfrak{iso}(3,1)$ one gives an equivalent $\sigma$ model, where the latter offers the natural choice of frame within its equivalence class, also on the field theory side. An automorphism that leaves the Poincar\'e patch, however, may a priori lead to an inequivalent model in the AdS/CFT context. I will not address this interesting point in more detail here -- it already applies to the algebraically equivalent abelian cases $p_i \wedge p_j$ and $k_i \wedge k_j$ for instance.} Namely, they follow by applying the TsT transformations associated to an $r$ matrix, directly to flat space. In the spirit of \cite{Seiberg:1999vs}, this gives an effective open string geometry corresponding to a field theory which is indeed noncommutative. In fact, its NC structure is encoded directly in the $r$ matrix.

\subsection*{TsT transformations in the open string picture}

To see the link between TsT transformations and the open string NC structure, consider the $O(D,D)$ formulation of TsT transformations (see e.g. \cite{Osten:2016dvf} in the present context). Under a TsT transformation with shift parameter $\gamma$, an arbitrary background $\mathcal{E}_{\mu\nu} = g_{\mu\nu} + B_{\mu\nu}$ with sufficient isometries transforms as
\begin{equation}
\mathcal{E} \rightarrow \tilde{\mathcal{E}} = \mathcal{E} \left(1+\Gamma \mathcal{E}\right)^{-1} = \tilde{g} + \tilde{B},
\end{equation}
where
\begin{equation}
\Gamma = - 2 \gamma \, r,
\end{equation}
with the abelian $r$ matrix $r$ realized in terms of differential operators as in equation \eqref{eq:rdiffop}, i.e. $\Gamma^{\mu\nu} = - 2 \gamma \, a^{\mu\nu}$. Of course, here $r$ matrices are just convenient bookkeeping devices to label TsT transformations \cite{vanTongeren:2015soa,Osten:2016dvf}. Following \cite{Schomerus:1999ug,Seiberg:1999vs}, see also \cite{Cornalba:2001sm}, the effective open string geometry attached to such a metric and $B$ field is
\begin{equation}
\label{eq:openstringparameters}
\begin{aligned}
G^{\mu\nu} & = \left(\frac{1}{\tilde{\mathcal{E}}}\right)_{S}^{\mu\nu},\\
\theta^{\mu\nu} & = 2 \pi \alpha^\prime  \left(\frac{1}{\tilde{\mathcal{E}}}\right)_{A}^{\mu\nu},
\end{aligned}
\end{equation}
where $S$ and $A$ denote the symmetric and antisymmetric part of the matrix respectively. In this open string picture the %transformed
background simplifies considerably, becoming
\begin{equation}
G^{\mu\nu} + \theta^{\mu\nu} = g^{\mu\nu} - 4 \pi \alpha^\prime \gamma\, a^{\mu\nu}.
\end{equation}
In other words, $\theta =  - 4 \pi \alpha^{\prime} \gamma\, r$. Doing a second TsT transformations simply adds the appropriate $r$ matrix term, and hence the above holds for a sequence of TsT transformations. These transformations are not required to commute. Of course it must be possible to find coordinates in the geometry such that the added TsT transformation can actually be performed. This is precisely the case with almost abelian $r$ matrices -- the TsT associated to $\hat{r}$ preserves the symmetries needed to do the TsT associated to $\bar{r}$, which both preserve the symmetries needed for $\tilde{r}$ in the rank six case. Hence at any stage in the sequence there is a choice of coordinates that permits the desired TsT transformation.\footnote{\label{footnote:AATsT=rdef} With this structure, the proof of equivalence between sequences of commuting TsT transformations and abelian $r$ matrices of \cite{Osten:2016dvf} immediately extends to the almost abelian case, providing a rigorous version of the arguments of \cite{Borsato:2016ose}. Moreover, dipole deformations -- also obtained via TsT transformations, now involving $\mathrm{S}^5$ or $\mathbb{R}^6$ respectively -- can be similarly viewed in terms of a twisted product \cite{Dasgupta:2001zu} associated to a noncommutativity parameter. Combining this with the spacetime noncommutativity parameter above then gives a combination of the two types of deformation. For compatible TsT transformations this noncommutativity parameter is again just the $r$ matrix, in line with the general twist picture.}

To get to a field theory from the open strings requires taking a low energy $\alpha^{\prime}\rightarrow 0$ limit. With $\tilde{\gamma} = \alpha^\prime \gamma$ fixed, this gives finite noncommutativity
\begin{equation}
\theta = - 4 \pi \tilde{\gamma} \,r,
\end{equation}
where\footnote{In general one should use the Kontsevich formula \cite{Kontsevich:1997vb,Cornalba:2001sm} here, giving higher order terms in agreement with the twisted product. This linear formula holds for rank four $r$ matrices in appropriate coordinates.}
\begin{equation}
[y^\mu \stackrel{\star}{,} y^\nu] = i \theta^{\mu\nu}.
\end{equation}
Note that $\theta$ need not be constant. There is one caveat in taking this limit: there are TsT setups where there are physical obstructions to taking this naive low energy limit, such as the would-be setup for canonical time-space NC SYM \cite{Seiberg:2000ms}, corresponding to a TsT transformation involving time (giving electric components in the $B$ field). There one gets four dimensional NC open string theory instead \cite{Seiberg:2000ms}. Its gravitational dual \cite{Gopakumar:2000na} is also different from the corresponding naive TsT transformation of $\ads$. Note that this caveat for instance applies to the ``analytic continuation'' of indices $(123)\rightarrow (012)$ that would take $r_2$ to the example of an $\mathfrak{r}_{3,-1}\oplus \mathbb{R}$ $r$ matrix considered in \cite{Borsato:2016ose}. Cases where time is explicitly involved, i.e. cases with electric $B$ field components, should hence be investigated on a case by case basis. This applies to $r_6$, $r_7$, $r_{10}$, $r_{12}$, $r_{13}$ and $r_{14}$ of \cite{Borsato:2016ose}. Null components are connected to spacelike ones by boosts and are ok, see e.g. \cite{Aharony:2000gz}. Modulo this caveat, applying commuting or noncommuting TsT transformations to flat space with D3 branes, gives a setup that limits to NC SYM, with its NC structure determined by the $r$ matrix. This matches the Drinfeld twist picture described above -- the factor of $2\pi$ is part of the difference between parameters in the closed string and field theory pictures respectively, as indicated below.

\subsection*{Closed string picture}

The closed string geometry follows by applying the TsT transformations to the brane geometry directly. This limits to the appropriate YB $\sigma$ model background by construction, as the TsT transformations only involve $\mathbb{R}^{1,3}$. Hence, abelian YB deformed strings are dual to NC versions of SYM, in line with the proposal of \cite{vanTongeren:2015uha}. Let me illustrate this general picture on the two examples above.

\subsection*{Examples -- $r_1$ -- $\mathfrak{h}_3 \oplus \mathbb{R}$}

In the conventions of \cite{Maldacena:1999mh}, the standard D3 brane metric is given by\footnote{The remaining supergravity fields do not affect our considerations. They are guaranteed to exist as all we are doing is T dualities and field redefinitions, in fact following immediately from them.}
\begin{align}
ds^2 & = \frac{1}{\sqrt{f}}(dx_\mu dx^\mu + \sqrt{f}(dr^2 + r^2 d\Omega_5^2),\nonumber\\
B & = 0, \qquad f(r)= 1 + (\alpha^\prime R^2)^2/r^4.
\end{align}
Applying transformations \eqref{eq:TsTr1} with $\gamma$ instead of $\eta$ gives
\begin{align}
ds^2 & = \frac{\gamma^2 x^- dx^- (2 dx^2 - x^- dx^-)  + f \left((dx^2)^2+ (dx^3)^2\right)}{\sqrt{f}(f+ \gamma^2)},\nonumber\\
& \quad - \frac{dx^+ dx^-}{\sqrt{f}} + \sqrt{f}(dr^2 + r^2 d\Omega_5^2),\\
B & = \frac{\gamma}{f+ \gamma^2} (- x^- dx^- \wedge dx^3 + dx^2 \wedge dx^3).\nonumber
\end{align}

For the near horizon low energy limit, take $r = \alpha^\prime R^2 /z$, $\tilde{\eta}= \gamma \alpha^\prime$ fixed, and $\alpha^\prime \rightarrow 0$, to find
\begin{align}
\frac{ds^2}{\alpha^\prime R^2}&=\frac{(dx^2)^2+(dx^3)^2+\tilde{\eta}^2 R^4 { x^-} d { x^-} (2 d { x^2}- { x^-}
   d { x^-})/z^4}{z^2+\tilde{\eta}^2 R^4/z^2}\nonumber\\
& \quad +\frac{-d { x^-} d { x^+}+dz^2}{z^2}+d\Omega_5^2,\nonumber\\
\frac{B}{\alpha^\prime R^2}& =\frac{\tilde{\eta} R^2   ( dx^2-{ x^-} d { x^-}) \wedge d x^3}{ \left(\tilde{\eta}^2 R^4+z^4\right)},
\end{align}
which, as expected, is precisely the background of the YB $\sigma$ model associated to $r_1$ as in eqn. \eqref{eq:backgroundr1}, with radius $\tilde{R} = \sqrt{\alpha^\prime} R$ reinstated, and $\eta = \tilde{\eta} R^2 = \tilde{\eta} \sqrt{\lambda}$, where $\lambda$ is the 't Hooft coupling. This background admits eight real supercharges.
 %in the conventions of e.g. \cite{Arutyunov:2009ga}.

To see the geometry the branes are placed in, consider instead the limit $r\rightarrow \infty$ to get
\begin{align}
ds^2 & = \frac{(dx^2)^2+ (dx^3)^2 + \gamma^2 x^- dx^-(2 dx^2 -x^- dx^-)}{1+ \gamma^2}\nonumber\\
&  \quad  - dx^+ dx^- + dy_k dy^k, \\
B & = \frac{\gamma }{1+ \gamma^2} (- x^- dx^- \wedge dx^3 + dx^2 \wedge dx^3),\nonumber
\end{align}
where the $y_k$, $k=1,\ldots,6$, are cartesian coordinates for $\mathbb{R}^6$. This is just the result of applying the above sequence of TsT transformations to flat space directly. In the low energy limit $\alpha^\prime \rightarrow 0$ with $\tilde{\gamma} = \gamma \alpha^\prime$ fixed, the effective open string geometry of equations \eqref{eq:openstringparameters} becomes\footnote{In this limit we effectively get $B \sim \gamma^{-1}$ and hence $\theta \sim 1/B$, as in \cite{Seiberg:1999vs}. Moreover, note that there is no critical value for $\gamma$ at which $B$ blows up, as there would be for canonical space-time noncommutativity \cite{Seiberg:2000ms}. There the $p_0 \wedge p_1$ type TsT transformation gives a $B$ field proportional to $(1- \gamma^2)^{-1}$.}
\begin{equation}
\theta = - 2 \pi \tilde{\gamma} (2 x^- \partial_{x^+} \wedge \partial_{x^3} + \partial_{x^2} \wedge \partial_{x^3}) = - 4 \pi \gamma r_1,
\end{equation}
and $G$ a flat ten dimensional metric. This corresponds to noncommutativity of the kind
\begin{equation}
[x^+ \stackrel{\star}{,} x^3] = - 4 \pi i \tilde{\gamma} x^-, \qquad \qquad [x^2 \stackrel{\star}{,} x^3]  = - 2 \pi i \tilde{\gamma},
\end{equation}
matching equations \eqref{eq:NC1} where we should use $\tilde{\eta}$ instead of $\eta$, with deformation parameters related by the effective string tension $T = \sqrt{\lambda}/2\pi$. This is the same relation as one finds for canonical NC SYM\footnote{Canonical NC SYM is contained in our present considerations. Concretely, rescaling $x^\mu \rightarrow b x^\mu$, $z \rightarrow b z$ in the present deformation of $\ads$ and considering the limit $b \rightarrow 0$ with $b^{-2} \eta = a^2$ constant gives the gravity dual of canonical NC SYM in the conventions of \cite{Maldacena:1999mh}. In this limit the commutator $[x^+ \stackrel{\star}{,} x^3]$ correspondingly vanishes, leaving only the standard $(x_2,x_3)$ noncommutativity. To make contact with \cite{Maldacena:1999mh} at the level of the brane geometry directly, rescale $x^2$ and $x^3$ by $\cos{\theta}$ and identify $\gamma = \tan{\theta}$.} or the $\beta$ deformation.

\subsection*{Examples -- $r_2$ -- $\mathfrak{r}_{3,-1}\oplus \mathbb{R}$}

The situation for the second example is completely analogous. Applying transformations \eqref{eq:TsTr2} to the brane background gives
\begin{align}
ds^2 & = \frac{f (d\rho^2 + \rho^2 d\xi^2 +(d x^3)^2) +\gamma^2 (\rho d \rho -d x^3)^2}{\sqrt{f}(f+  \gamma^2(1+\rho^2))},\nonumber\\
& \quad  -\frac{1}{\sqrt{f}} (dx^0)^2 + \sqrt{f}(dr^2 + r^2 d\Omega_5^2),\\
B & = \frac{\gamma }{f+  \gamma^2(1+\rho^2)} (- \rho d\rho \wedge d\xi - \rho^2 d\xi \wedge dx^3).\nonumber
\end{align}
which as above limits to the associated YB background of equations \eqref{eq:backgroundr2}, admitting no supersymmetry. The open string picture now gives
\begin{align}
\theta &= 2 \pi \tilde{\gamma} (\rho^{-1} \partial_\rho \wedge \partial_\xi + \partial_\xi \wedge \partial_{x^3}) \\
&= 2 \pi \tilde{\gamma} (\partial_{x^1} \wedge \partial_{x^2} - (x^1 \partial_{x^2} \wedge \partial_{x^3} - x^2 \partial_{x^1} \wedge \partial_{x^3})),\nonumber
\end{align}
matching equations \eqref{eq:NC2}. This is just a combination of canonical $(x_1,x_2)$ NC SYM and its $(\xi,x^3)$ analogue of \cite{Hashimoto:2005hy}, in both the field theory and string pictures, matching the TsT structure. Similar explicit constructions can be readily pursued for the other $r$ matrices listed in \cite{Borsato:2016ose}.

\subsection*{Remarks regarding supersymmetry}

The brane picture discussed above should be stable to provide a notion of duality. This would be guaranteed by supersymmetry of the backgrounds. There are unimodular deformations that preserve a quarter of the original superstring, such as the first example above. Others do not preserve any supersymmetry, however, and here the proposed duality may well break down due to quantum effects. The $\gamma_i$ deformation \cite{Frolov:2005dj}, a three parameter generalization of the $\beta$ deformation that breaks all supersymmetry, is an illustrative example. Here conformal symmetry is broken even in the planar limit of the field theory \cite{Fokken:2013aea}. Correspondingly, without supersymmetry certain string modes are expected to become tachyonic and lead to a deformation of $\mathrm{AdS}_5$. Similar subtleties presumably affect (some of) the nonsupersymmetric cases here as well. Even in these cases some notion of duality may nevertheless remain -- despite lacking a clean AdS/CFT picture, for the $\gamma_i$ deformation spectra can be matched for a large class of states in the planar limit \cite{Fokken:2014soa,vanTongeren:2013gva}.

%%%%%%%%%%%%%%%%%%%%%%%%%%%%%%%%%%%%%%
\section{Concluding remarks}\label{sec:conclusions}
%%%%%%%%%%%%%%%%%%%%%%%%%%%%%%%%%%%%%%

I showed that most almost abelian YB deformations of the $\ads$ superstring are AdS/CFT duals of NC versions of SYM, where the corresponding noncommutativity parameter is the $r$ matrix, matching the general picture of \cite{vanTongeren:2015uha}. These deformations of $\ads$ can be realized via noncommuting sequences of noncommuting TsT transformations. The corresponding dual NC structure is a combination of the NC structures associated to the individual TsT transformations, mirroring the $r$ matrix picture.

There are some almost abelian deformation that involve generators of the conformal algebra that do not have a natural action in the brane geometry. It is important to understand whether and if so, how, one can find similar brane constructions for them. The same applies to their abelian building blocks already. This has been done for the abelian twist of SYM on $\mathbb{R} \times \mathrm{S}^3$ built out of the Cartan generators of $\mathfrak{so}(4) \subset \mathfrak{so}(4,2)$ \cite{Dhokarh:2008ki}. Beyond almost abelian deformations, there are nonabelian unimodular deformations that cannot be represented as sequences of TsT transformations, instead requiring nonabelian T duality. It would be interesting to provide explicit open string pictures for the NC structures expected to be associated to these, and, at the algebraic level, to construct the associated twists. In this case the open string noncommutativity should be equal to the $r$ matrix as well. It is also worth mentioning that general YB models can be viewed as adding a $B$ field equal to the \emph{inverse} of the $r$ matrix in a nonabelian T dual picture \cite{Borsato:2016pas} which appears closely related to the present picture for almost abelian models, where a $B$ field equal to the $r$ matrix is added to the ``inverse'' geometry. The latter applies both before and after the near horizon limit, and it would be interesting to see how the former carries through the brane construction. Moreover, it is important to understand whether in the Poincar\'{e} patch there is a physical distinction between $r$ matrices related by inner automorphisms that leave this patch. Clarifying the meaning of generic unimodular YB models with electric $B$ field in string theory and AdS/CFT is also relevant.

In broader terms, it would be great to classify the possible unimodular deformations of $\ads$ for the full superalgebra $\mathfrak{psu}(2,2|4)$, and investigate their AdS/CFT duals in detail. In particular, it would be nice to understand conclusively whether an inhomogeneous but unimodular $r$ matrix exists. One might hope that at least an $r$ matrix exists that becomes unimodular in a contraction limit \cite{Pachol:2015mfa}, as there is a natural string candidate there \cite{Arutyunov:2014cra,Arutyunov:2014jfa}.\footnote{There is in fact a jordanian deformation of $\ads$, i.e. not a string, which contracts to a jordanian deformation of flat space \cite{Hoare:2016hwh} (see also \cite{Borowiec:2015wua}) that does solve supergravity \cite{Hoare:2016hwh}.} Nonunimodular models may prove worth further investigation as well, as they are formally T dual to string models \cite{Hoare:2015wia,Arutyunov:2015mqj,Hoare:2016wsk} and it would be interesting to see what they correspond to. Finally, while the spectrum of the $\eta$ model can be found (assuming formal light-cone gauge fixing) \cite{Arutynov:2014ota}, it remains an important open question to understand homogeneous deformed models, beyond those based on the Cartan subalgebra, see e.g. \cite{vanTongeren:2013gva}, at the quantum level.

%%%%%%%%%%%%%%%%%%%%%%%%%%%%%%%%%%%%%%
\section*{Acknowledgments}
%%%%%%%%%%%%%%%%%%%%%%%%%%%%%%%%%%%%%%

I would like to thank Riccardo Borsato, Ben Hoare, and Linus Wulff for discussions, and Arkady Tseytlin, Riccardo Borsato, Ben Hoare, and Linus Wulff for comments on the draft. I am supported by L.T. I  acknowledge further support by the SFB project 647  ``Space-Time-Matter: Analytic and Geometric Structures'' and the European Union Programme FP7/2007-2013/ under REA Grant Agreement No 317089.

\appendix

\section{The conformal algebra}
\label{app:confalg}

The four dimensional conformal algebra, $\mathfrak{so}(4,2)$, can be represented as
\begin{equation*}
\begin{aligned}
{\color{white} ?} [m_{\mu\nu}, p_\rho] &  = \eta_{\nu\rho} p_\mu - \eta_{\mu\rho} p_\nu, \hspace{14pt} [m_{\mu\nu}, k_\rho]  = \eta_{\nu\rho} k_\mu- \eta_{\mu\rho} k_\nu,\\
[m_{\mu\nu}, D] & = 0, \hspace{14pt} [D,p_\mu]=p_\mu, \hspace{14pt} [D,K_\mu]=-K_\mu,\\
[p_\mu,k_\nu] & =2 m_{\mu\nu} + 2 \eta_{\mu\nu} D, \\
[m_{\mu\nu},m_{\rho\sigma}] & = \eta_{\mu\rho} m_{\nu\sigma} + \mbox{perms.}
\end{aligned}
\end{equation*}
These antihermitian generators can be realized as differential operators on $\mathbb{R}^{1,3}$ as
\begin{align}
p_\mu & = \partial_\mu, \hspace{20pt} k_\mu  = x_\alpha x^\alpha \partial_\mu -2 x_\mu x^\nu \partial_\nu \nonumber \\
m_{\mu\nu} & = x_\mu \partial_\nu - x_\nu \partial_\mu, \hspace{20pt} D = - x^\mu \partial_\mu,
\end{align}
where the $p$ generate translations, the $m$ rotations and boosts, and the $k$ special conformal transformations, or for instance in the fundamental representation of $\mathfrak{su}(2,2) \simeq \mathfrak{so}(4,2)$ as
\begin{equation}
\begin{aligned}
p_\mu & = \tfrac{1}{2}(\gamma_\mu - \gamma_\mu \gamma_4), &  k_\mu & = \tfrac{1}{2}(\gamma_\mu + \gamma_\mu \gamma_4),\\
m_{\mu \nu} & = \tfrac{1}{2} \gamma_\mu \gamma_\nu, & D & = \tfrac{1}{2} \gamma_4.
\end{aligned}
\end{equation}
Here the $\gamma^i$ are $4\times4$ $\gamma$ matrices, for instance
\begin{equation}
\begin{aligned}
&\gamma^0 = i \sigma_3 \otimes \sigma_0,  &\gamma^1 = \sigma_2 \otimes \sigma_2, &&\gamma^2 = -\sigma_2 \otimes \sigma_1, \\ &\gamma^3 = \sigma_1 \otimes \sigma_0, &\gamma^4 = \sigma_2 \otimes \sigma_3, &&\gamma^5 = -i \gamma^0,
\end{aligned}
\end{equation}
where $\sigma_0 = 1_{2\times2}$ and the remaining $\sigma_i$ are the Pauli matrices.

\section{Drinfeld twists and $r$ matrices}
\label{app:Drinfeldtwist}

Consider the standard Hopf algebra associated to $\mathcal{U}(\mathfrak{g})$, the universal enveloping algebra of a semisimple Lie algebra $\mathfrak{g}$, with coproduct $\Delta$, counit $\epsilon$ and antipode $s$. A Drinfeld twist $F$ is an invertible element of $\mathcal{U}(\mathfrak{g}) \otimes \mathcal{U}(\mathfrak{g})$ which satisfies the cocycle condition \cite{DrinfeldTwistRef,Reshetikhin:1990ep}
\begin{equation}
(F \otimes 1)(\Delta \otimes 1) F = (1 \otimes F)(1\otimes \Delta) F,
\end{equation}
and normalization condition
\begin{equation}
(\epsilon \otimes 1) F =(1\otimes \epsilon) F = 1 \otimes 1.
\end{equation}
Moreover, for $F$ to represent a deformation,
\begin{equation}
\label{genFexpansion}
F = 1\otimes 1 + i \eta F^{(1)} + \mathcal{O}(\eta^2),
\end{equation}
where $\eta$ is a deformation parameter.\footnote{Here I flipped the inconsequential sign of the deformation parameter with respect to the $\sigma$ model twist in the conventions of \cite{vanTongeren:2015uha}, to have the structures in the main text match directly.} Let us express $F$ as a sum of terms in $\mathcal{U}(\mathfrak{g})\otimes \mathcal{U}(\mathfrak{g})$
\begin{equation}
F = f^\beta \otimes f_\beta, \hspace{10pt} F^{-1} = \bar{f}^\beta \otimes \bar{f}_\beta,
\end{equation}
where $f^\beta$, $f_\beta$, $\bar{f}^\beta$, and $\bar{f}_\beta$ denote in principle distinct elements of $\mathcal{U}(\mathfrak{g})$, and we have an implicit (infinite) sum over $\beta$. $F$ can now be used to deform (twist) the Hopf algebra by changing the original coproduct and antipode $s$ to
\begin{equation}
\label{eq:twistedhopfalgebra}
\begin{aligned}
\Delta_F (X) & = F \Delta(X) F^{-1},\\
s_F(X) & = f^\alpha s(f_\alpha) s(X) s(\bar{f}^\beta) \bar{f}_\beta.
\end{aligned}
\end{equation}
The cocycle condition on $F$ guarantees coassociativity of the twisted coproduct, as well as associativity of the $\star$ product used in the main text.

A classical $r$ matrix for $\mathfrak{g}$ is an $r \in \mathfrak{g} \otimes \mathfrak{g}$ that solves the classical Yang-Baxter equation
\begin{equation}
\label{eq:CYBEMAT}
[r_{12},r_{13}]+[r_{12},r_{23}]+[r_{13},r_{23}] = 0.
\end{equation}
Here $r_{mn}$ denotes the matrix realization of $r$ acting in spaces $m$ and $n$ in a tensor product. Classical $r$ matrices are in one to one correspondence with Drinfeld twists in the following sense \cite{DrinfeldTwistRef} (see also e.g. \cite{Giaquinto:1994jx}). First, the classical $r$ matrix
\begin{equation}
r_{12} = \tfrac{1}{2}(\mathcal{F}^{(1)}_{12} - \mathcal{F}^{(1)}_{21}),
\end{equation}
solves the CYBE. Second, twists that have the same classical $r$ matrix give equivalent deformations of the algebra. Third, a twist exists for any solution of the CYBE, though a general explicit construction is not known.

\subsection*{Almost abelian twists}

Given an abelian $r$ matrix such as $\hat{r}$ one can define an associated abelian twist $\hat{F}$ as
\begin{equation}
\hat{F} = e^{i \eta \hat{r}},
\end{equation}
which is readily verified to satisfy the required properties. Due to the special structure of almost abelian $r$ matrices one can then define a twist for $r = \hat{r} + \bar{r}$ as
\begin{equation}
F = \bar{F} \hat{F}.
\end{equation}
This construction works due identities like $\mbox{ad}_{\bar{r}_{13} + \bar{r}_{23}} \, \hat{r}_{12} =0$ which hold thanks to the possible defining commutation relations (\ref{eq:comm1}-\ref{eq:comm3}). For the rank six case of the main text, $F = \tilde{F} \bar{F} \hat{F}$.

\bibliographystyle{elsarticle-num}

\bibliography{stijnsbibfile}

\begin{thebibliography}{10}
\expandafter\ifx\csname url\endcsname\relax
  \def\url#1{\texttt{#1}}\fi
\expandafter\ifx\csname urlprefix\endcsname\relax\def\urlprefix{URL }\fi
\expandafter\ifx\csname href\endcsname\relax
  \def\href#1#2{#2} \def\path#1{#1}\fi

\bibitem{Maldacena:1997re}
J.~M. Maldacena, {The Large N limit of superconformal field theories and
  supergravity}, Int. J. Theor. Phys. 38 (1999) 1113--1133, [Adv. Theor. Math.
  Phys.2,231(1998)].
\newblock \href {http://arxiv.org/abs/hep-th/9711200}
  {\path{arXiv:hep-th/9711200}}, \href
  {http://dx.doi.org/10.1023/A:1026654312961}
  {\path{doi:10.1023/A:1026654312961}}.

\bibitem{Arutyunov:2009ga}
G.~Arutyunov, S.~Frolov, {Foundations of the $\ads$ Superstring. Part I},
  J.Phys. A42 (2009) 254003.
\newblock \href {http://arxiv.org/abs/0901.4937} {\path{arXiv:0901.4937}},
  \href {http://dx.doi.org/10.1088/1751-8113/42/25/254003}
  {\path{doi:10.1088/1751-8113/42/25/254003}}.

\bibitem{Beisert:2010jr}
N.~Beisert, C.~Ahn, L.~F. Alday, Z.~Bajnok, J.~M. Drummond, et~al., {Review of
  AdS/CFT Integrability: An Overview}, Lett.Math.Phys. 99 (2012) 3--32.
\newblock \href {http://arxiv.org/abs/1012.3982} {\path{arXiv:1012.3982}},
  \href {http://dx.doi.org/10.1007/s11005-011-0529-2}
  {\path{doi:10.1007/s11005-011-0529-2}}.

\bibitem{Bombardelli:2016rwb}
D.~Bombardelli, A.~Cagnazzo, R.~Frassek, F.~Levkovich-Maslyuk, F.~Loebbert,
  S.~Negro, I.~M. Sz\'{e}cs\'{e}nyi, A.~Sfondrini, S.~J. van Tongeren,
  A.~Torrielli, An integrability primer for the gauge-gravity correspondence:
  an introduction, J.Phys.A 49 (2016) 320301.
\newblock \href {http://arxiv.org/abs/1606.02945} {\path{arXiv:1606.02945}},
  \href {http://dx.doi.org/10.1088/1751-8113/49/32/320301}
  {\path{doi:10.1088/1751-8113/49/32/320301}}.

\bibitem{Lunin:2005jy}
O.~Lunin, J.~M. Maldacena, {Deforming field theories with $U(1)\times U(1)$
  global symmetry and their gravity duals}, JHEP 0505 (2005) 033.
\newblock \href {http://arxiv.org/abs/hep-th/0502086}
  {\path{arXiv:hep-th/0502086}}, \href
  {http://dx.doi.org/10.1088/1126-6708/2005/05/033}
  {\path{doi:10.1088/1126-6708/2005/05/033}}.

\bibitem{Frolov:2005ty}
S.~Frolov, R.~Roiban, A.~A. Tseytlin, {Gauge-string duality for superconformal
  deformations of N=4 super Yang-Mills theory}, JHEP 0507 (2005) 045.
\newblock \href {http://arxiv.org/abs/hep-th/0503192}
  {\path{arXiv:hep-th/0503192}}, \href
  {http://dx.doi.org/10.1088/1126-6708/2005/07/045}
  {\path{doi:10.1088/1126-6708/2005/07/045}}.

\bibitem{Frolov:2005dj}
S.~Frolov, {Lax pair for strings in Lunin-Maldacena background}, JHEP 0505
  (2005) 069.
\newblock \href {http://arxiv.org/abs/hep-th/0503201}
  {\path{arXiv:hep-th/0503201}}, \href
  {http://dx.doi.org/10.1088/1126-6708/2005/05/069}
  {\path{doi:10.1088/1126-6708/2005/05/069}}.

\bibitem{Klimcik:2002zj}
C.~Klimcik, {Yang-Baxter sigma models and dS/AdS T duality}, JHEP 0212 (2002)
  051.
\newblock \href {http://arxiv.org/abs/hep-th/0210095}
  {\path{arXiv:hep-th/0210095}}, \href
  {http://dx.doi.org/10.1088/1126-6708/2002/12/051}
  {\path{doi:10.1088/1126-6708/2002/12/051}}.

\bibitem{Klimcik:2008eq}
C.~Klimcik, {On integrability of the Yang-Baxter sigma-model}, J.Math.Phys. 50
  (2009) 043508.
\newblock \href {http://arxiv.org/abs/0802.3518} {\path{arXiv:0802.3518}},
  \href {http://dx.doi.org/10.1063/1.3116242} {\path{doi:10.1063/1.3116242}}.

\bibitem{Delduc:2013qra}
F.~Delduc, M.~Magro, B.~Vicedo, {An integrable deformation of the $\ads$
  superstring action}, Phys.Rev.Lett. 112 (2014) 051601.
\newblock \href {http://arxiv.org/abs/1309.5850} {\path{arXiv:1309.5850}},
  \href {http://dx.doi.org/10.1103/PhysRevLett.112.051601}
  {\path{doi:10.1103/PhysRevLett.112.051601}}.

\bibitem{Sfetsos:2013wia}
K.~Sfetsos, {Integrable interpolations: From exact CFTs to non-Abelian
  T-duals}, Nucl.Phys. B880 (2014) 225--246.
\newblock \href {http://arxiv.org/abs/1312.4560} {\path{arXiv:1312.4560}},
  \href {http://dx.doi.org/10.1016/j.nuclphysb.2014.01.004}
  {\path{doi:10.1016/j.nuclphysb.2014.01.004}}.

\bibitem{Hollowood:2014qma}
T.~J. Hollowood, J.~L. Miramontes, D.~M. Schmidtt, {An Integrable Deformation
  of the $\ads$ Superstring}, J.Phys. A47~(49) (2014) 495402.
\newblock \href {http://arxiv.org/abs/1409.1538} {\path{arXiv:1409.1538}},
  \href {http://dx.doi.org/10.1088/1751-8113/47/49/495402}
  {\path{doi:10.1088/1751-8113/47/49/495402}}.

\bibitem{Demulder:2015lva}
S.~Demulder, K.~Sfetsos, D.~C. Thompson, {Integrable $\lambda$-deformations:
  Squashing Coset CFTs and $AdS_5\times S^5$}, JHEP 07 (2015) 019.
\newblock \href {http://arxiv.org/abs/1504.02781} {\path{arXiv:1504.02781}},
  \href {http://dx.doi.org/10.1007/JHEP07(2015)019}
  {\path{doi:10.1007/JHEP07(2015)019}}.

\bibitem{Delduc:2014kha}
F.~Delduc, M.~Magro, B.~Vicedo, {Derivation of the action and symmetries of the
  $q$-deformed $\ads$ superstring}, JHEP 1410 (2014) 132.
\newblock \href {http://arxiv.org/abs/1406.6286} {\path{arXiv:1406.6286}},
  \href {http://dx.doi.org/10.1007/JHEP10(2014)132}
  {\path{doi:10.1007/JHEP10(2014)132}}.

\bibitem{Kawaguchi:2014qwa}
I.~Kawaguchi, T.~Matsumoto, K.~Yoshida, {Jordanian deformations of the $\ads$
  superstring}, JHEP 1404 (2014) 153.
\newblock \href {http://arxiv.org/abs/1401.4855} {\path{arXiv:1401.4855}},
  \href {http://dx.doi.org/10.1007/JHEP04(2014)153}
  {\path{doi:10.1007/JHEP04(2014)153}}.

\bibitem{vanTongeren:2015uha}
S.~J. van Tongeren, {Yang-Baxter deformations, AdS/CFT, and
  twist-noncommutative gauge theory}, Nucl. Phys. B904 (2016) 148--175.
\newblock \href {http://arxiv.org/abs/1506.01023} {\path{arXiv:1506.01023}},
  \href {http://dx.doi.org/10.1016/j.nuclphysb.2016.01.012}
  {\path{doi:10.1016/j.nuclphysb.2016.01.012}}.

\bibitem{Borsato:2016ose}
R.~Borsato, L.~Wulff, {Target space supergeometry of $\eta$ and
  $\lambda$-deformed strings}, JHEP 10 (2016) 045.
\newblock \href {http://arxiv.org/abs/1608.03570} {\path{arXiv:1608.03570}},
  \href {http://dx.doi.org/10.1007/JHEP10(2016)045}
  {\path{doi:10.1007/JHEP10(2016)045}}.

\bibitem{Arutyunov:2015qva}
G.~Arutyunov, R.~Borsato, S.~Frolov, {Puzzles of $\eta$-deformed AdS$_5 \times$
  S$^5$}, JHEP 12 (2015) 049.
\newblock \href {http://arxiv.org/abs/1507.04239} {\path{arXiv:1507.04239}},
  \href {http://dx.doi.org/10.1007/JHEP12(2015)049}
  {\path{doi:10.1007/JHEP12(2015)049}}.

\bibitem{Hoare:2016ibq}
B.~Hoare, S.~J. van Tongeren, {Non-split and split deformations of AdS$_5$}, J.
  Phys. A49~(48) (2016) 484003.
\newblock \href {http://arxiv.org/abs/1605.03552} {\path{arXiv:1605.03552}},
  \href {http://dx.doi.org/10.1088/1751-8113/49/48/484003}
  {\path{doi:10.1088/1751-8113/49/48/484003}}.

\bibitem{vanTongeren:2015soa}
S.~J. van Tongeren, {On classical Yang-Baxter based deformations of the $\ads$
  superstring}, JHEP 06 (2015) 048.
\newblock \href {http://arxiv.org/abs/1504.05516} {\path{arXiv:1504.05516}},
  \href {http://dx.doi.org/10.1007/JHEP06(2015)048}
  {\path{doi:10.1007/JHEP06(2015)048}}.

\bibitem{Osten:2016dvf}
D.~Osten, S.~J. van Tongeren, {Abelian Yang-Baxter Deformations and TsT
  transformations}\href {http://arxiv.org/abs/1608.08504}
  {\path{arXiv:1608.08504}}.

\bibitem{Kyono:2016jqy}
H.~Kyono, K.~Yoshida, {Supercoset construction of Yang-Baxter deformed
  AdS$_5\times$S$^5$ backgrounds}, PTEP B03 (2016) 083.
\newblock \href {http://arxiv.org/abs/1605.02519} {\path{arXiv:1605.02519}},
  \href {http://dx.doi.org/10.1093/ptep/ptw111}
  {\path{doi:10.1093/ptep/ptw111}}.

\bibitem{Hoare:2016hwh}
B.~Hoare, S.~J. van Tongeren, {On jordanian deformations of AdS$_5$ and
  supergravity}, J. Phys. A49~(43) (2016) 434006.
\newblock \href {http://arxiv.org/abs/1605.03554} {\path{arXiv:1605.03554}},
  \href {http://dx.doi.org/10.1088/1751-8113/49/43/434006}
  {\path{doi:10.1088/1751-8113/49/43/434006}}.

\bibitem{Orlando:2016qqu}
D.~Orlando, S.~Reffert, J.-i. Sakamoto, K.~Yoshida, {Generalized type IIB
  supergravity equations and non-Abelian classical r-matrices}, J. Phys.
  A49~(44) (2016) 445403.
\newblock \href {http://arxiv.org/abs/1607.00795} {\path{arXiv:1607.00795}},
  \href {http://dx.doi.org/10.1088/1751-8113/49/44/445403}
  {\path{doi:10.1088/1751-8113/49/44/445403}}.

\bibitem{Arutyunov:2015mqj}
G.~Arutyunov, S.~Frolov, B.~Hoare, R.~Roiban, A.~A. Tseytlin, {Scale invariance
  of the $\eta$-deformed $AdS_5\times S^5$ superstring, T-duality and modified
  type II equations}, Nucl. Phys. B903 (2016) 262--303.
\newblock \href {http://arxiv.org/abs/1511.05795} {\path{arXiv:1511.05795}},
  \href {http://dx.doi.org/10.1016/j.nuclphysb.2015.12.012}
  {\path{doi:10.1016/j.nuclphysb.2015.12.012}}.

\bibitem{Wulff:2016tju}
L.~Wulff, A.~A. Tseytlin, {Kappa-symmetry of superstring sigma model and
  generalized 10d supergravity equations}, JHEP 06 (2016) 174.
\newblock \href {http://arxiv.org/abs/1605.04884} {\path{arXiv:1605.04884}},
  \href {http://dx.doi.org/10.1007/JHEP06(2016)174}
  {\path{doi:10.1007/JHEP06(2016)174}}.

\bibitem{Hoare:2016wsk}
B.~Hoare, A.~A. Tseytlin, {Homogeneous Yang-Baxter deformations as non-abelian
  duals of the AdS$_5$ sigma-model}, J. Phys. A49~(49) (2016) 494001.
\newblock \href {http://arxiv.org/abs/1609.02550} {\path{arXiv:1609.02550}},
  \href {http://dx.doi.org/10.1088/1751-8113/49/49/494001}
  {\path{doi:10.1088/1751-8113/49/49/494001}}.

\bibitem{Borsato:2016pas}
R.~Borsato, L.~Wulff, {Integrable deformations of T-dual $\sigma$ models}\href
  {http://arxiv.org/abs/1609.09834} {\path{arXiv:1609.09834}}.

\bibitem{Hashimoto:1999ut}
A.~Hashimoto, N.~Itzhaki, {Noncommutative Yang-Mills and the AdS / CFT
  correspondence}, Phys.Lett. B465 (1999) 142--147.
\newblock \href {http://arxiv.org/abs/hep-th/9907166}
  {\path{arXiv:hep-th/9907166}}, \href
  {http://dx.doi.org/10.1016/S0370-2693(99)01037-0}
  {\path{doi:10.1016/S0370-2693(99)01037-0}}.

\bibitem{Maldacena:1999mh}
J.~M. Maldacena, J.~G. Russo, {Large N limit of noncommutative gauge theories},
  JHEP 9909 (1999) 025.
\newblock \href {http://arxiv.org/abs/hep-th/9908134}
  {\path{arXiv:hep-th/9908134}}, \href
  {http://dx.doi.org/10.1088/1126-6708/1999/09/025}
  {\path{doi:10.1088/1126-6708/1999/09/025}}.

\bibitem{Matsumoto:2014nra}
T.~Matsumoto, K.~Yoshida, {Lunin-Maldacena backgrounds from the classical
  Yang-Baxter equation - towards the gravity/CYBE correspondence}, JHEP 1406
  (2014) 135.
\newblock \href {http://arxiv.org/abs/1404.1838} {\path{arXiv:1404.1838}},
  \href {http://dx.doi.org/10.1007/JHEP06(2014)135}
  {\path{doi:10.1007/JHEP06(2014)135}}.

\bibitem{Matsumoto:2014gwa}
T.~Matsumoto, K.~Yoshida, {Integrability of classical strings dual for
  noncommutative gauge theories}, JHEP 1406 (2014) 163.
\newblock \href {http://arxiv.org/abs/1404.3657} {\path{arXiv:1404.3657}},
  \href {http://dx.doi.org/10.1007/JHEP06(2014)163}
  {\path{doi:10.1007/JHEP06(2014)163}}.

\bibitem{Schomerus:1999ug}
V.~Schomerus, {D-branes and deformation quantization}, JHEP 9906 (1999) 030.
\newblock \href {http://arxiv.org/abs/hep-th/9903205}
  {\path{arXiv:hep-th/9903205}}, \href
  {http://dx.doi.org/10.1088/1126-6708/1999/06/030}
  {\path{doi:10.1088/1126-6708/1999/06/030}}.

\bibitem{Seiberg:1999vs}
N.~Seiberg, E.~Witten, {String theory and noncommutative geometry}, JHEP 9909
  (1999) 032.
\newblock \href {http://arxiv.org/abs/hep-th/9908142}
  {\path{arXiv:hep-th/9908142}}, \href
  {http://dx.doi.org/10.1088/1126-6708/1999/09/032}
  {\path{doi:10.1088/1126-6708/1999/09/032}}.

\bibitem{Seiberg:2000ms}
N.~Seiberg, L.~Susskind, N.~Toumbas, {Strings in background electric field,
  space / time noncommutativity and a new noncritical string theory}, JHEP 0006
  (2000) 021.
\newblock \href {http://arxiv.org/abs/hep-th/0005040}
  {\path{arXiv:hep-th/0005040}}, \href
  {http://dx.doi.org/10.1088/1126-6708/2000/06/021}
  {\path{doi:10.1088/1126-6708/2000/06/021}}.

\bibitem{Dasgupta:2001zu}
K.~Dasgupta, M.~Sheikh-Jabbari, {Noncommutative dipole field theories}, JHEP
  0202 (2002) 002.
\newblock \href {http://arxiv.org/abs/hep-th/0112064}
  {\path{arXiv:hep-th/0112064}}, \href
  {http://dx.doi.org/10.1088/1126-6708/2002/02/002}
  {\path{doi:10.1088/1126-6708/2002/02/002}}.

\bibitem{Metsaev:1998it}
R.~Metsaev, A.~A. Tseytlin, {Type IIB superstring action in AdS(5) x S**5
  background}, Nucl.Phys. B533 (1998) 109--126.
\newblock \href {http://arxiv.org/abs/hep-th/9805028}
  {\path{arXiv:hep-th/9805028}}, \href
  {http://dx.doi.org/10.1016/S0550-3213(98)00570-7}
  {\path{doi:10.1016/S0550-3213(98)00570-7}}.

\bibitem{Ovando:2006}
G.~Ovando, {Four dimensional symplectic Lie algebras}, Beitrage zur Algebra und
  Geometry 47 (2006) 419, \url{https://eudml.org/doc/226718}.

\bibitem{Borowiec:2008se}
A.~Borowiec, J.~Lukierski, V.~N. Tolstoy, {New twisted quantum deformations of
  D=4 super-Poincare algebra} (2008) 205--216\href
  {http://arxiv.org/abs/0803.4167} {\path{arXiv:0803.4167}}.

\bibitem{Szabo:2006wx}
R.~J. Szabo, {Symmetry, gravity and noncommutativity}, Class.Quant.Grav. 23
  (2006) R199--R242.
\newblock \href {http://arxiv.org/abs/hep-th/0606233}
  {\path{arXiv:hep-th/0606233}}, \href
  {http://dx.doi.org/10.1088/0264-9381/23/22/R01}
  {\path{doi:10.1088/0264-9381/23/22/R01}}.

\bibitem{Dimitrijevic:2014dxa}
M.~Dimitrijevic, L.~Jonke, A.~Pacho\l{}, {Gauge Theory on Twisted
  $\kappa$-Minkowski: Old Problems and Possible Solutions}, SIGMA 10 (2014)
  063.
\newblock \href {http://arxiv.org/abs/1403.1857} {\path{arXiv:1403.1857}},
  \href {http://dx.doi.org/10.3842/SIGMA.2014.063}
  {\path{doi:10.3842/SIGMA.2014.063}}.

\bibitem{Cornalba:2001sm}
L.~Cornalba, R.~Schiappa, {Nonassociative star product deformations for D-brane
  world volumes in curved backgrounds}, Commun.Math.Phys. 225 (2002) 33--66.
\newblock \href {http://arxiv.org/abs/hep-th/0101219}
  {\path{arXiv:hep-th/0101219}}, \href
  {http://dx.doi.org/10.1007/s002201000569} {\path{doi:10.1007/s002201000569}}.

\bibitem{Kontsevich:1997vb}
M.~Kontsevich, {Deformation quantization of Poisson manifolds. 1.}, Lett. Math.
  Phys. 66 (2003) 157--216.
\newblock \href {http://arxiv.org/abs/q-alg/9709040}
  {\path{arXiv:q-alg/9709040}}, \href
  {http://dx.doi.org/10.1023/B:MATH.0000027508.00421.bf}
  {\path{doi:10.1023/B:MATH.0000027508.00421.bf}}.

\bibitem{Gopakumar:2000na}
R.~Gopakumar, J.~M. Maldacena, S.~Minwalla, A.~Strominger, {S duality and
  noncommutative gauge theory}, JHEP 0006 (2000) 036.
\newblock \href {http://arxiv.org/abs/hep-th/0005048}
  {\path{arXiv:hep-th/0005048}}, \href
  {http://dx.doi.org/10.1088/1126-6708/2000/06/036}
  {\path{doi:10.1088/1126-6708/2000/06/036}}.

\bibitem{Aharony:2000gz}
O.~Aharony, J.~Gomis, T.~Mehen, {On theories with lightlike noncommutativity},
  JHEP 0009 (2000) 023.
\newblock \href {http://arxiv.org/abs/hep-th/0006236}
  {\path{arXiv:hep-th/0006236}}, \href
  {http://dx.doi.org/10.1088/1126-6708/2000/09/023}
  {\path{doi:10.1088/1126-6708/2000/09/023}}.

\bibitem{Hashimoto:2005hy}
A.~Hashimoto, K.~Thomas, {Non-commutative gauge theory on d-branes in melvin
  universes}, JHEP 0601 (2006) 083.
\newblock \href {http://arxiv.org/abs/hep-th/0511197}
  {\path{arXiv:hep-th/0511197}}, \href
  {http://dx.doi.org/10.1088/1126-6708/2006/01/083}
  {\path{doi:10.1088/1126-6708/2006/01/083}}.

\bibitem{Fokken:2013aea}
J.~Fokken, C.~Sieg, M.~Wilhelm, {Non-conformality of ${{\gamma }_{i}}$-deformed
  N = 4 SYM theory}, J.Phys. A47 (2014) 455401.
\newblock \href {http://arxiv.org/abs/1308.4420} {\path{arXiv:1308.4420}},
  \href {http://dx.doi.org/10.1088/1751-8113/47/45/455401}
  {\path{doi:10.1088/1751-8113/47/45/455401}}.

\bibitem{Fokken:2014soa}
J.~Fokken, C.~Sieg, M.~Wilhelm, {A piece of cake: the ground-state energies in
  $\gamma_{i}$ -deformed $ \mathcal{N} $ = 4 SYM theory at leading wrapping
  order}, JHEP 1409 (2014) 78.
\newblock \href {http://arxiv.org/abs/1405.6712} {\path{arXiv:1405.6712}},
  \href {http://dx.doi.org/10.1007/JHEP09(2014)078}
  {\path{doi:10.1007/JHEP09(2014)078}}.

\bibitem{vanTongeren:2013gva}
S.~J. van Tongeren, {Integrability of the $\ads$ superstring and its
  deformations}, J.Phys. A47~(43) (2014) 433001.
\newblock \href {http://arxiv.org/abs/1310.4854} {\path{arXiv:1310.4854}},
  \href {http://dx.doi.org/10.1088/1751-8113/47/43/433001}
  {\path{doi:10.1088/1751-8113/47/43/433001}}.

\bibitem{Dhokarh:2008ki}
D.~Dhokarh, S.~S. Haque, A.~Hashimoto, {Melvin Twists of global AdS(5) x S(5)
  and their Non-Commutative Field Theory Dual}, JHEP 08 (2008) 084.
\newblock \href {http://arxiv.org/abs/0801.3812} {\path{arXiv:0801.3812}},
  \href {http://dx.doi.org/10.1088/1126-6708/2008/08/084}
  {\path{doi:10.1088/1126-6708/2008/08/084}}.

\bibitem{Pachol:2015mfa}
A.~Pacho{\l}, S.~J. van Tongeren, {Quantum deformations of the flat space
  superstring}, Phys. Rev. D93 (2016) 026008.
\newblock \href {http://arxiv.org/abs/1510.02389} {\path{arXiv:1510.02389}},
  \href {http://dx.doi.org/10.1103/PhysRevD.93.026008}
  {\path{doi:10.1103/PhysRevD.93.026008}}.

\bibitem{Arutyunov:2014cra}
G.~Arutyunov, S.~J. van Tongeren, {The $\mathrm{AdS}_5 \times \mathrm{S}^5$
  mirror model as a string}, Phys.Rev.Lett. 113 (2014) 261605.
\newblock \href {http://arxiv.org/abs/1406.2304} {\path{arXiv:1406.2304}},
  \href {http://dx.doi.org/10.1103/PhysRevLett.113.261605}
  {\path{doi:10.1103/PhysRevLett.113.261605}}.

\bibitem{Arutyunov:2014jfa}
G.~Arutyunov, S.~J. van Tongeren, {Double Wick rotating Green-Schwarz strings},
  JHEP 1505 (2015) 027.
\newblock \href {http://arxiv.org/abs/1412.5137} {\path{arXiv:1412.5137}},
  \href {http://dx.doi.org/10.1007/JHEP05(2015)027}
  {\path{doi:10.1007/JHEP05(2015)027}}.

\bibitem{Borowiec:2015wua}
A.~Borowiec, H.~Kyono, J.~Lukierski, J.-i. Sakamoto, K.~Yoshida, {Yang-Baxter
  sigma models and Lax pairs arising from $\kappa$-Poincar\'e $r$-matrices},
  JHEP 04 (2016) 079.
\newblock \href {http://arxiv.org/abs/1510.03083} {\path{arXiv:1510.03083}},
  \href {http://dx.doi.org/10.1007/JHEP04(2016)079}
  {\path{doi:10.1007/JHEP04(2016)079}}.

\bibitem{Hoare:2015wia}
B.~Hoare, A.~A. Tseytlin, {Type IIB supergravity solution for the T-dual of the
  $\eta$-deformed AdS$_{5} \times$ S$^{5}$ superstring}, JHEP 10 (2015) 060.
\newblock \href {http://arxiv.org/abs/1508.01150} {\path{arXiv:1508.01150}},
  \href {http://dx.doi.org/10.1007/JHEP10(2015)060}
  {\path{doi:10.1007/JHEP10(2015)060}}.

\bibitem{Arutynov:2014ota}
G.~Arutyunov, M.~de~Leeuw, S.~J. van Tongeren, {The exact spectrum and mirror
  duality of the $(\ads)_\eta$ superstring}, Theor.Math.Phys. 182~(1) (2015)
  23--51.
\newblock \href {http://arxiv.org/abs/1403.6104} {\path{arXiv:1403.6104}},
  \href {http://dx.doi.org/10.1007/s11232-015-0243-9}
  {\path{doi:10.1007/s11232-015-0243-9}}.

\bibitem{DrinfeldTwistRef}
V.~Drinfeld, {On constant quasi-classical solutions of the Yang-Baxter quantum
  equation}, Sov. Math. Dokl. 28 (1983) 667.

\bibitem{Reshetikhin:1990ep}
N.~Reshetikhin, {Multiparameter quantum groups and twisted quasitriangular Hopf
  algebras}, Lett.Math.Phys. 20 (1990) 331--335.
\newblock \href {http://dx.doi.org/10.1007/BF00626530}
  {\path{doi:10.1007/BF00626530}}.

\bibitem{Giaquinto:1994jx}
A.~Giaquinto, J.~J. Zhang, {Bialgebra actions, twists, and universal
  deformation formulas}, J.Pure Appl.Algebra 128 (1998) 133--151.
\newblock \href {http://arxiv.org/abs/hep-th/9411140}
  {\path{arXiv:hep-th/9411140}}, \href
  {http://dx.doi.org/10.1016/S0022-4049(97)00041-8}
  {\path{doi:10.1016/S0022-4049(97)00041-8}}.

\end{thebibliography}

%%%%%%%%%%%%%%%%%%%%%%%%%%%%%%%%%%%%%%

%%%%%%%%%%%%%%%%%%%%%%%%%%%%%%%%%%%%%%
\end{document}